\documentstyle[12pt]{article}
\def\a{\alpha}
\def\l{\lambda}

\def\ket#1{|{#1}\rangle}  % |#1>
\def\bra#1{\langle{#1}|}  % <#1|
\def\norm#1#2{\langle{#1}|{#2}\rangle}   % <#1|#2>

\input epsf

\begin{document} 
%\preprint{IMSc-98/04/15}
\title{Chern-Simons Theory, Knot Invariants, \\Vertex Models
and Three-manifold Invariants{\footnote{Invited talk at the
Workshop on {\it Frontiers of Field Theory, Quantum Gravity and
Strings,} 12th to 21st December 1996, Puri, India.}}}
\author{Romesh K. Kaul\\
The Institute of Mathematical Sciences, \\
 Taramani, Madras 600113, India.\\
{\it email:~kaul@imsc.ernet.in} \\
{\it IMSc-98/04/15}}
\maketitle

\vskip1.0cm

\begin{abstract}
Chern-Simons theories, which are topological quantum
field theories, provide a field theoretic
framework for the study of knots and links in three
dimensions.  These  are rare examples of
quantum field theories which can be exactly and
explicitly solved. Expectation values of Wilson link 
operators yield a class of link invariants, the simplest
of them is the famous Jones polynomial. Other
invariants are more powerful than that of Jones.
These new invariants are sensitive to  the chirality of all
knots at least upto ten crossing number unlike those
of Jones which are blind to the  chirality of some of them.
However, all these invariants are still not
good  enough to distinguish  a class of knots
called mutants. These link invariants can be alternately obtained from two dimensional
vertex models. The $R$-matrix of such  a model in a 
particular limit of the spectral parameter provides a 
representation  of the braid group.  This  in turn is 
used to construct the link invariants. Exploiting  
theorems of Lickorish and Wallace and also  those of Kirby, Fenn
and Rourke  which relate three-manifolds to surgeries on 
framed links, these link invariants in $S^3$ can also 
be used to construct three-manifold invariants.  
\end{abstract}

\vskip0.5cm

\section{Introduction} Topological field theories provide a bridge between quantum
field theories and topology of low dimensional manifolds.
A topological field theory is independent of the metric~
$g_{\mu \nu}$~ of the manifold on which it is defined. This
means that expectation value of the energy-momentum tensor, which is given by the
functional variation of the partition function with respect
to the metric~ $g_{\mu \nu}$~ is zero, $ <T^{\mu \nu}>~ =~
{\delta Z \over {\delta g_{\mu \nu}}} ~=~ 0$. The topological
operators $W$ of such a theory are metric independent, 
${\delta \over\delta g_{\mu \nu}} <W> ~=~0$.

An example of a topological field theory is Chern-Simons
gauge theory on a three-manifold. This theory provides a field
theoretic frame work for the study of knots and links in a
given three manifold\cite{ati}$^-$\cite{kau4} .  
It was Schwarz  who first conjectured \cite{sch} that the now
famous Jones polynomial \cite{jon} may be related to
Chern-Simons theory. Witten in his pioneering
paper\cite{wit} set up the general framework to study knots
and links through Chern-Simons field theories.
Wilson loop operators are the topological operators of this
theory. Expectation value of these operators are the
topological invariants for knots and links. The simplest of
these invariants is that of Jones which is associated with
spin half representation in an $SU(2)$ Chern-Simons 
theory\cite{wit} .
Other representations and other semi-simple gauge groups
yield other knot invariants. These invariants are also 
intimately related to the integrable vertex models in 
two dimensions\cite{wad,tur} . Representation theory of
quantum groups provided yet another  direct framework in which
these invariants can be studied\cite{kiril} .
A mathematically important development is that these
link invariants provide a method of obtaining topological
invariants for three-manifold\cite{res}$^-$\cite{lic} . 
In the following, we shall review these
developments.

\vskip0.5cm

\section{Chern-Simons field theory and link invariants} For a matrix valued 
connection one-form $A$ of the gauge group $G$, the
Chern-Simons action $S$ is given by
\begin{equation}
kS ~=~ {\frac k {4\pi}} \int_{M^3} tr(AdA~+~{\frac{2}{3}} A^3)
\label{ac}
\end{equation}
\noindent The coupling constant $k$ takes integer values in the 
quantum theory. We shall, except when stated otherwise, take 
the gauge group $G$ to be $SU(2)$ and the three-manifold 
$M^3$ to be $S^3$ for definiteness.
Clearly action (\ref{ac}) does not have any metric of $M^3$ in it.
The topological operators are the Wilson loop (knot)
operators defined as
\begin{equation} 
W_j [C]~=~ tr_j Pexp \oint_C A_j  \label{wil}
\end{equation}  
\noindent for an oriented knot $C$ carrying spin $j$
representation. A few simple knots are:

\vskip0.2cm
\centerline{\epsfbox{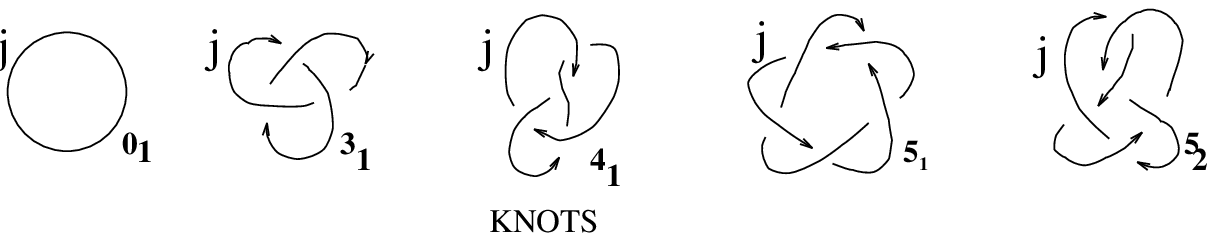}}
\vskip0.2cm

\noindent The Wilson operators are independent of the metric of the
three-manifold. For a link $L$
made up of oriented component knots $C_1, C_2, \ldots C_r$
carrying spin $j_1,
j_2,\ldots j_r$ representations respectively, we have the
Wilson link operator
defined as
\begin{equation}
W_{j_1j_2\ldots j_r} [L] \ = \ \prod_{\ell=1}^r \ W_{j_\ell}
[C_\ell] \label{wil1}
\end{equation}
\noindent A few two-component links are:

\centerline{\epsfbox{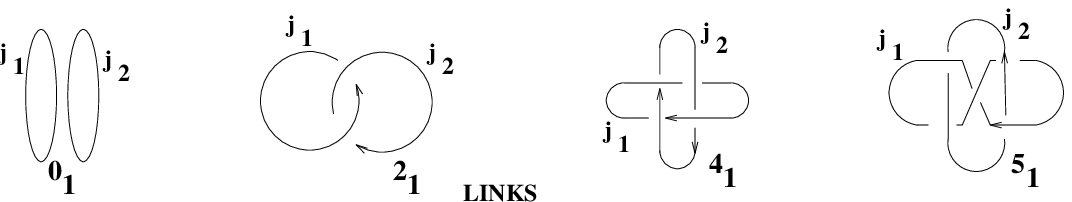}}
\noindent We are interested in the functional averages of
the link operators:
\begin{equation} 
V_{j_1j_2\ldots j_r}[L] \ = \ Z^{-1} \int_{S^3} [dA]
W_{j_1j_2\ldots j_r} [L]
e^{ikS}, ~~~~~~~{\rm where}~~ Z \ = \ \int_{S^3} [dA] e^{ikS} \label{jon}    
\end{equation}
\noindent Here the integrands in the functional integrals are metric 
independent. So is the measure \cite{kauraj}. Therefore, these
expectation values depend
only on the isotopy type of the oriented link $L$ and the
set of
representations $j_1, j_2\ldots j_r$ associated with the
component knots.

The expectation values of Wilson link operators (\ref{jon}) can be 
determined exactly in Chern-Simons theory. For this purpose
two ingredients, one from quantum field theory and other
from mathematics of braids, are used \cite{kau2}:
\vskip0.2cm
 
{\it (i) Field theoretic input:} The first ingredient is that
the Chern-Simons theory
on a three-manifold with boundary is essentially
characterized by a corresponding Wess-Zumino conformal
field theory on that boundary\cite{wit}: 
\vskip0.2cm
\centerline{\epsfbox{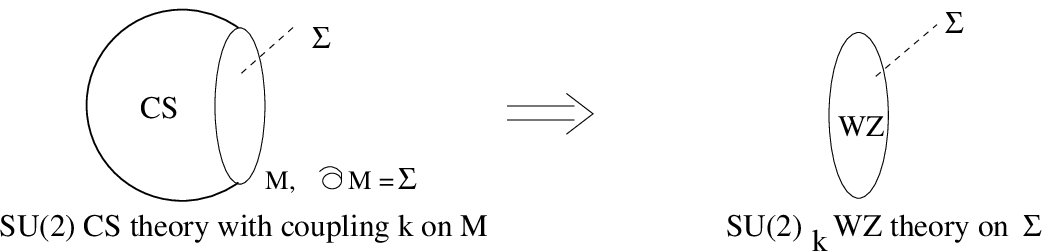}} 
\vskip0.2cm
\noindent And Chern-Simons functional average for Wilson
lines ending at $n$ points in the boundary is described by the
associated Wess-Zumino theory on the boundary with $n$
punctures carrying the representations of the free Wilson
lines:
\vskip0.2cm
\centerline{\epsfbox{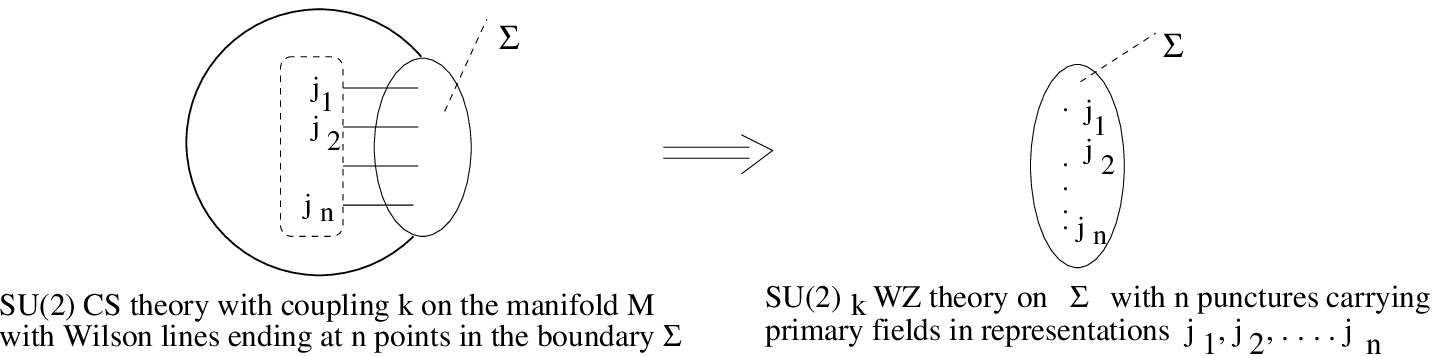}} 
\vskip0.2cm
\noindent The Chern-Simons functional integral can be
represented \cite{wit} by a vector in the Hilbert space $\cal{H}$
associated with the space of $n$-point correlator of the Wess-Zumino
conformal field theory on the boundary $\Sigma$. 
In fact, these correlators provide a basis
for this boundary Hilbert space. There are more
than one possible
basis. These different bases are related by duality of the
correlators of the
conformal field theory\cite{tsu}. 

\vskip0.2cm

{\it (ii) Mathematical input:} The second input we shall need is the close 
connection knots and links have with braids. 
An $n$-braid is a collection of non-intersecting strands
connecting $n$ points
on a horizontal plane to $n$ points on another horizontal
plane directly below
the first set of $n$ points. The strands are not allowed to
go back upwards at
any point in their travel. The braid may be projected onto a
plane with the two
horizontal planes collapsing to two parallel rigid rods. The
over-crossings and
under-crossings of the strands are to be clearly marked.
When all the strands
are identical, we have ordinary braids. 
The theory of such
braids, first developed by Artin\cite{art} , is well studied.
These braids form a group. However, for our purpose here we need
to orient the
individual strands and further distinguish them by putting
different colours on
them. We shall represent different colours by different
$SU(2)$ spins. These braids, unlike braids made from
unoriented identical strands, have a more general structure than a
group. These instead
form a groupoid.  The necessary aspects of
the theory of such
braids have been  presented in ref.(8).

One way of relating the braids to knots and links is through closure of
braids. We obtain the closure of a braid by connecting the ends of the
first, second, third, .... strands from above to the ends of the 
respective first, second, third ..... strands  from below as shown in
(A): 
\vskip0.2cm
\centerline{\epsfbox{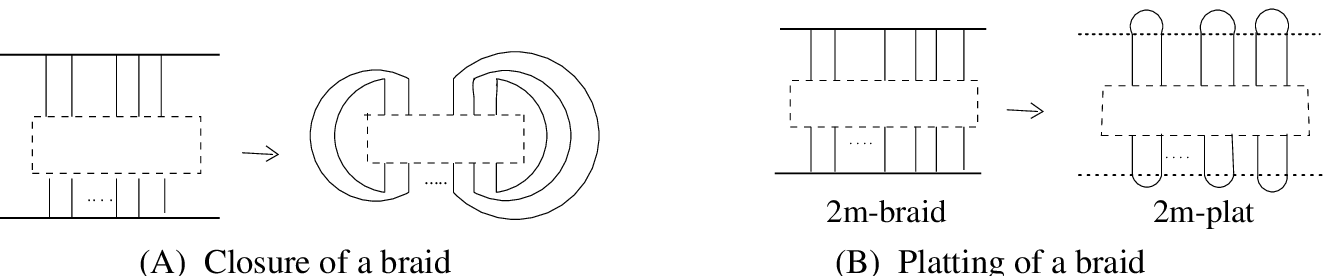}}
\vskip0.2cm
\noindent There is a theorem by Alexander\cite{alex} which states that {\it any knot or link can be
obtained as closure of a braid.} This construction of a knot or link is
not unique.

There is another construction associated with braids which relates
them to
knots and links.  This is called platting. Consider a
$2m$-braid, with pairwise adjacent strands carrying the same
colour and
opposite orientations.  Then
connect the $(2i-1)$th strand with $(2i)$th from above as
well as
below.  This yields the plat of the given braid as shown in (B) above. 
Then there is a theorem due to Birman\cite{bir} which relates
plats to links. This states that {\it a coloured-oriented link can be represented (though not
uniquely) by
the plat of an oriented-coloured $2m$-braid}.

Use of these two inputs, namely relation of Chern-Simons theory to
the boundary Wess-Zumino theory and presentation of knots and links as
closures or platts of  braids leads to an explicit and complete solution
of the Chern-Simons theory.
For this purpose, consider a manifold $S^3$         
from which two non-intersecting three-balls are removed.
This manifold                  
has two boundaries, each an $S^2$.  We place $2n$  Wilson line-integrals 
over lines connecting these two boundaries  through a weaving 
pattern $\bf B$ as  shown in the Figure (a) below:
\vskip0.2cm
\centerline{\epsfbox{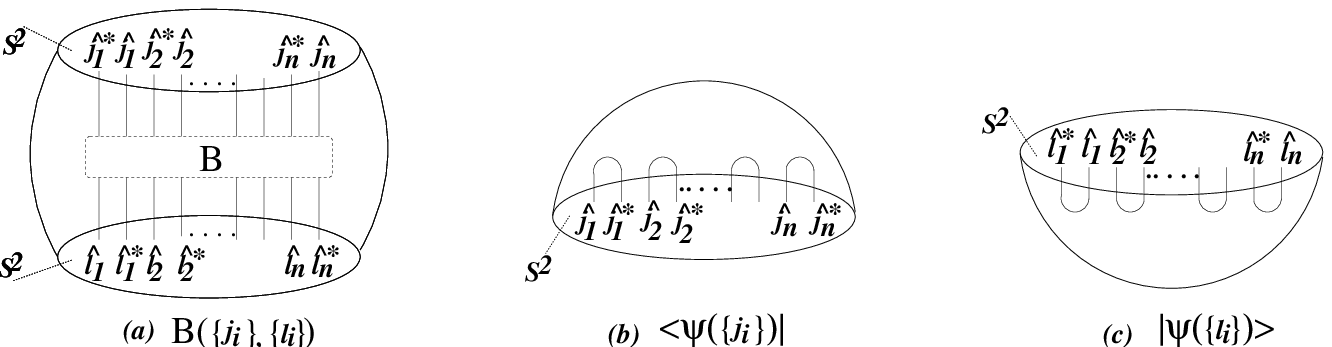}} 
\vskip0.2cm
\noindent This is a $2n-$braid placed in this manifold.  
The strands are specified  on the upper boundary by giving $2n$ 
assignments $(\hat{j}^*_1,~
\hat{j}_1,~ \hat{j}^*_2, ~\hat{j}_2, ..... ~\hat{j}^*_n, ~\hat{j}_n)$. 
Here $\hat{j}~ =~ (j,~ \epsilon)$  represents the spin $j$ and 
orientation $\epsilon~ (\epsilon = \pm 1$  for a  strand going into 
or away from the boundary) and conjugate assignment $\hat{j}^*~
 =~(j, ~ -\epsilon)$ indicates reversal of the orientation.
Similar specifications are done with respect to the lower boundary by
the spin-orientation assignments $(\hat{\ell}_1, ~\hat{\ell}^*_1,~
\hat{\ell}_2, ~ \hat{\ell}^*_2, .... ~\hat{\ell}_n, ~\hat{\ell}^*_n)$.
 Then the assignments $\{ \hat{\ell}_i \}$ are just a permutation of
$\{ \hat{j}_i^\ast \}$. Chern-Simons functional integral over this manifold 
is a state in the tensor product of the Hilbert spaces
associated with the two boundaries, ${\cal{H}}_1 \, \otimes
{\cal{H}}_2$. This state can be expanded in terms of some convenient
basis\cite{kau2} .  These bases are given by the conformal blocks for 
$2n$-point correlators of the  $SU(2)_k$ Wess-Zumino conformal 
field theory. 

An arbitrary braid can be generated by a sequence of
elementary braidings. The eigen values  of these elementary
braids are given by conformal field theory.
A knot is given with some framing. Standard framing is such
that the self-linking number of the knot (linking number 
of the knot and its frame) is zero. The braiding eigenvalues depend on the 
framing. For standard framing  eigen-values for  an
elementary braiding  two strands carrying spins $j,~j'$ and with
orientations $\epsilon, ~\epsilon'$ are\cite{kau2} :
$$
\begin{array}{lc}
\lambda_t (\hat{j}, \hat{j}') & =  \left\{
\begin{array}{lcl}
                                 \lambda^{(+)}_t (j,j') & =
&
(-)^{j+j'-t} \, q^{(C_j+C_{j'})/2 + C_{min(j,j')} - C_t/2}
\,~~~~~~ {\rm if} \,
\epsilon 
\epsilon' = +
1 \\
& \\
                                 (\lambda^{(-)}_t (j,j'))^{-1} & =
&
(-)^{|j-j'|-t} \, q^{|C_j-C_{j'}|/2 - C_t/2}
\,~~~~~~~~~~~~~~~~~ {\rm
if} \, \epsilon \epsilon' = - 1
\end{array}  \right.
\end{array}    
$$
\noindent where $q~=~ exp( \frac {2\pi i} { k+2})$ and $C_j$ is the
quadratic Casimir invariant of the spin $j$ representation, $C_j~=~j(j+1)$ and
$t$ takes the values allowed in the product of representations of
spin $j$ and $j'$ by the fusion rules of $SU(2)_k$
Wess-Zumino conformal field theory, $t~=$ $~|j-j'|,~|j-j'|+1,~......
~{\rm min}(j + j', ~k - j -j' )$. When $\epsilon \epsilon' = +1$, the
two strands have the same orientation and the braid generator
introduces a right-handed half-twist. On the other hand for $\epsilon
\epsilon' = -1$, two strands are anti-parallel and braid generator
introduces a left-handed half-twist. Thus $\lambda^{(+)}_t (j,j') $
and $\lambda^{(-)}_t (j,j')$ are the eigen-values for elementary braidings 
introducing  right-handed half-twists in
parallely and anti-parallely oriented strands respectively.

Writing the weaving pattern  ${\bf B}$
in Figure (a) above in terms of the elementary braids,  the Chern-Simons
functional integral over this manifold is given by a matrix
$ {\bf B} (\{j_i\}, \{ \ell_i \}) $ in ${\cal{H}}_1 \, \otimes
{\cal{H}}_2$.

To plat this braid, we consider two balls with Wilson lines as
shown in Figures (b) and (c) above. We glue  these respectively 
from above and below onto the manifold of Figure (a).  This 
yields a link in $S^3$. 

The Chern-Simons functional integral over the ball (c) is given by  
a vector in the Hilbert space associated with the $S^2$ boundary.
This vector  ~$\vert \psi (\{\ell_i \})\rangle$ ~can again be 
written in terms of a convenient basis of this Hilbert space.
Similarly, the functional integral over the ball of Figure (b)
above is a vector ~$\langle \psi (\{j_i \}) \vert$~ in the associated
dual Hilbert space. Gluing these two balls on to each other gives $n$
disjoint unknots carrying spins $j_1,~j_2,~....~j_n$  in $S^3$. Their
invariant factorizes into the invariants for $n$ individual unknots. Thus 
the inner product of the vectors representing the functional integrals
over manifolds (b) and (c) is given by 
$$
\langle \psi(\{ j_i \}) \vert  \psi (\{ j_i \})\rangle~=~
\prod^n_{i~=~1} ~[2j_i + 1]
$$
\noindent where $~[2j + 1]~$ is the invariant for an unknot carrying
spin $j$ and the square brackets represent the $q-$numbers:
$[x]~=~(q^{\frac x 2} -q^{-\frac x 2})/(q^{\frac 1 2}- q^{-\frac 1
2})$.

Next gluing the two balls (b) and (c) on to the manifold of Figure
(a), is just taking the matrix element of the matrix ${\bf B}$ 
between these two vectors. This would thus yield a link-invariant \cite{kau2}:
\vskip0.2cm

{\bf Proposition:}~{\it Expectation value of a Wilson
operator for an arbitrary} $2n$ {\it coloured-oriented link with a
plat
representation in terms of a coloured-oriented braid}
${\bf {B}}(\{ j_i \}, \{ \ell_i \})$ {\it generated as a word in 
terms of the braid generators is given by}
$$
V[L] \, = \, \langle \psi(\{ j_i \}) \vert ~{\bf {B}}(\{ j_i \},
\{ \ell_i \}) ~\vert \psi (\{ \ell_i \})\rangle
$$

This proposition along with the earlier stated
result of
Birman, allows us to evaluate these invariants for any
arbitrary link. Jones polynomials are obtained by placing spin $1/2$
representations on all the component knots. When spin $1$
representations are placed on them, we have  Akutsu-Wadati/Kauffman
invariant\cite{kauf,wad}.  For higher spins, we have new invariants.
 
These invariants are generally sensitive to the chirality of many knots. 
For example, for left- and right-handed  trefoils $T_L, ~ T_R$, these
invariants are different: $V_j (T_L) \neq V_j(T_R) $ 
 for $j = {\frac 1 2},~ 1,~ {\frac 3 2} ~....$
The invariants for the mirror reflected knots are give by
simple complex conjugation. These invariants do not detect
chirality of all knots.
For example, upto ten crossing numbers, there are
six chiral knots, $9_{42}$,  $10_{48}$,  $10_{71}$, $10_{91}$, 
$10_{104}$  and $10_{125}$ (as listed in the knot
tables of Rolfsen's book\cite{rol}) which are not
distinguished from their mirror images by
Jones (spin 1/2) polynomials:  
\vskip0.2cm
\centerline{\epsfbox{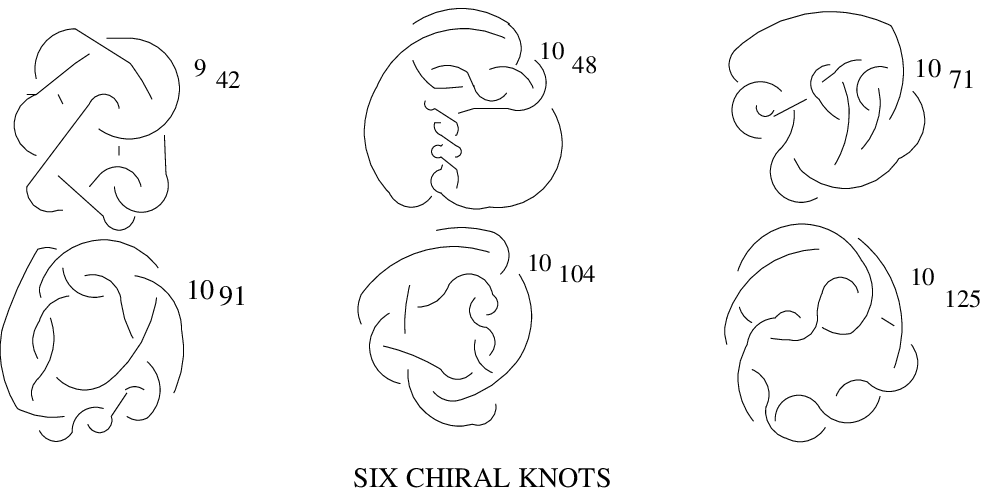}}
\vskip0.2cm

\noindent Kauffman/Akutsu-Wadati (spin
one) polynomials             
do detect the chirality of four of them, namely $10_{48},~
10_{91},
~10_{104} \,~ {\rm and}\, ~ 10_{125}$.  But for $9_{42} \, ~{\rm
and} \,~
10_{71}$       
both Jones and Kauffman polynomials are not  changed
under chirality                  
transformation ${(q \rightarrow q^{-1})}$.  However,  the
new spin
$3/2$ invariants are powerful enough to distinguish these
knots
from their mirror images\cite{kau3} .

While it does appear that these new invariants are more
powerful as we go up in  spin, $j = 1/2, 1, 3/2
\ldots$, it is not true that chirality of all knots can be 
detected by these invariants. We shall give below
an example where none of the invariants obtained from
Chern-Simons theories detect the chirality of a $16$ crossing
knot. 

\vskip0.5cm

\section{Mutants and their Chern-Simons invariants}

Besides chirality, there is another interesting property of  
knots and links which we would like to be detected by knot
invariants. This is the ``mutation'' of knots and links.
To study this, consider  a link $L_1$ obtained from two rooms
~\epsfbox{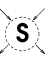}~ and ~\epsfbox{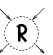}~ with two 
strands going in and two leaving in each of them
as shown in the Figure $L_1$:

\vskip0.2cm
\centerline{\epsfbox{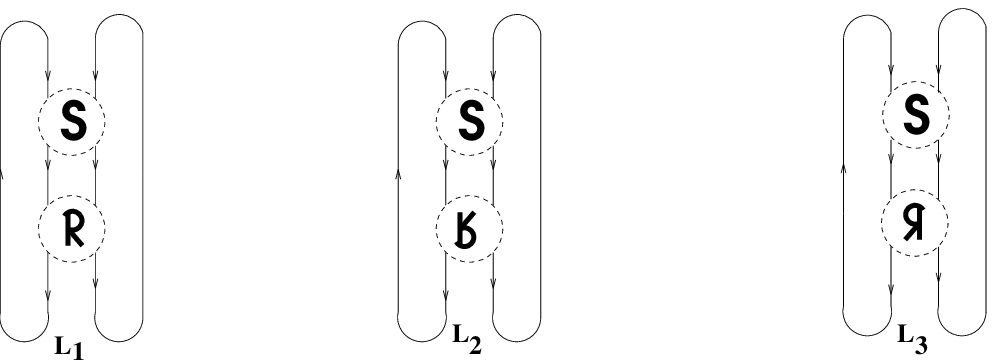}}
\vskip0.2cm

\noindent The mutant links are obtained in the following way:
(i) Remove one of the rooms, say ~\epsfbox{Sec3fig2.eps}~ from $L_1$
 and rotate it through $\pi$
about any one of the three orthogonal axes ($\gamma_i$) as
shown in the Figure:

\centerline{\epsfbox{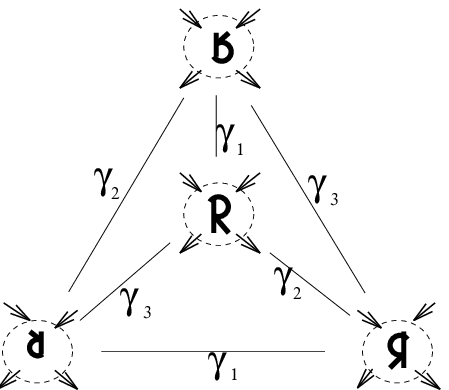}}
\vskip0.2cm

\noindent  Clearly only two of these rotations are
independent:
$\gamma_3~=~\gamma_1 * \gamma_2.$
(ii) Change the orientations of the lines
inside the rotated room ~\epsfbox{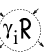}~ to
match with the fixed orientations of the external
legs of the original
room~ \epsfbox{Sec3fig2.eps}.~
(iii) Then, replace this room
back in $L_1$. This yields  for $\gamma_1$ and $\gamma_2$ mutations
of ~\epsfbox{Sec3fig2.eps} ~mutant links
$L_2$ and $L_3$ as shown above.

We shall argue that no invariants obtained from Chern-Simons theory 
distinguish these mutants\cite{kau4} . In the following discussion in this Section we shall
take the gauge group $G$ of the Chern-Simons theory to be arbitrary
 and shall not specialize to the $SU(2)$  gauge group. 

Next observe that the link $L_1$ in $S^3$ can be obtained by
gluing a three-ball
containing room ~\epsfbox{Sec3fig1.eps}~ as shown in Figure (i) below
 with another three-ball with oppositely oriented boundary $S^2$
containing room ~\epsfbox{Sec3fig2.eps}~ as shown in Figure (ii). 
Similarly, gluing Figures (iiia) and (iva) on to Figure (i) will 
give corresponding mutant links $L_2$ and $L_3$ respectively: 

\vskip0.2cm
\centerline{\epsfbox{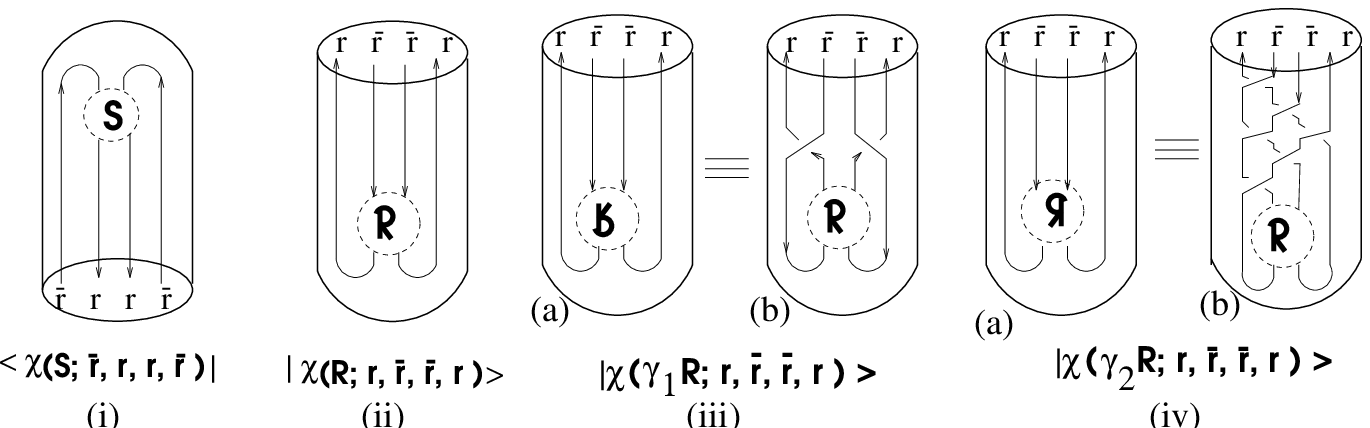}}
\vskip0.2cm

\noindent Notice, the diagrams (a) and (b) in each of the Figures (iii)
and (iv) are equivalent; these can be changed into each other by simple
isotopic moves of the strands. The functional integrals over these balls
can be represented by vectors in the Hilbert space associated
with the four-punctured boundary $S^2$ of each of them. For example, 
for the three-ball in Figure (ii), we may write the functional integral in 
a convenient basis~ $ {\ket {\phi^{side}_l(r,{\bar r},{\bar r},r)}} $~
characterized by the four-point correlators of the
associated  Wess-Zumino conformal field theory on $S^2$ as:
$$
{\ket {\chi(R;~r,{\bar r},{\bar r},r)}}~=~\sum_l \nu_l(R) ~
{\ket {\phi^{side}_l(r,{\bar r},{\bar r},r)}} 
$$
\noindent where $\nu_l(R)$ are coefficients characterizing the entanglements
in the room   ~\epsfbox{Sec3fig2.eps}~  and the basis with superscript
 ``side" refer to  an eigen-basis for
the elementary braiding generators $b_1$ and $b_3$ introducing left-handed 
half-twists in the first two and last two anti-parallel strands respectively:
\begin{eqnarray}
b_1{\ket {\phi_l^{side}(r,\bar r,\bar r,r)}}&~=~&(\l_l^{(-)}(r,\bar r))^{-1}
{\ket {\phi_l^{side}(\bar r, r,\bar r,r)}}\nonumber\\
b_3 {\ket {\phi_l^{side}(r,\bar r,\bar r,r)}}&~=~
&(\l_l^{(-)}(r,\bar r))^{-1}{\ket {\phi_l^{side}(r,\bar r,r, \bar r)}}
\label{eva}
\end{eqnarray}
\noindent Here the twisted side two strands are 
antiparallel and carry representations $r$ and $\bar r$, the index $l$
runs over all the irreducible representations in the fusion rule of
$r\otimes \bar r$
of the corresponding Wess-Zumino model. An equivalent basis
$\ket {\phi_m^{cent}}$ is one where braid generator $b_2$,
which introduces right-handed half-twists in the central two strands, is
diagonal: 
\begin{equation}
b_2{\ket {\phi_m^{cent}(\bar r,r,r,\bar r)}}~ =~
\l_m^{(+)}(r,r)~\ket {\phi_m^{cent}(\bar r,r,r,\bar r)}~. \nonumber
\end{equation}
Since this  refers to twisting of parallel strands both carrying representation $r$, 
the index $m$ refers to the allowed irreducible representations in the 
fusion rule
$r \otimes r$ of the corresponding Wess-Zumino model. 
The two bases are related by
$q$-Racah coefficient of the quantum group 
$G_q$\cite{kiril,kau1,kau2}:
$$
{\ket {\phi_l^{side}(\bar r,r,r,\bar r)}}~=~\sum_m
a_{lm}\left
[\matrix {\bar r&r\cr r&\bar r}\right ]
{\ket {\phi_m^{cent}(\bar r,r,r,\bar r)}}~.
$$
\noindent The eigenvalues
$\l_l^{(-)}(r,\bar r)$ and $\l_m^{(+)}(r,r)$ for right-handed
half-twists in two anti-parallel and
parallel strands are respectively \cite {kau1}:
\begin{equation}
\l_l^{(-)}(r,\bar r)~=~(-1)^{\epsilon}~q^{C_l/2}~;~
\l_m^{(+)}(r,r)~=~(-1)^{\epsilon}~q^{2C_r - C_m/2}~, \label{eva1}
\end{equation}
where $C_r$, $C_m$ and $C_l$ are the quadratic Casimir invariants in the
representations
$r$, $m$ and $l$ respectively. Depending upon the representation $l$
($m$)
occurring symmetrically or antisymmetrically in the
tensor product $r\otimes \bar r$ ($r\otimes r$), $\epsilon=\pm1$.
Further  $q~=~\exp {2\pi i/ (k+C_v)}$, where
$C_v$ is the  quadratic Casimir invariant in the adjoint representation and $k$
is
the Chern-Simons coupling.

Now notice that the diagram (iiib) can be generated by applying the braid
generators $b_1$ and $b_3^{-1}$ on the diagram  in Figure (ii) with
interchanged orientation and representation assignments on the first and second,
third and fourth strands ending on the boundary. Therefore, we can 
relate  the vectors representing the fuctional integral over these manifolds
as
$$
{\ket { \chi(\gamma_1 R;~r,\bar r,\bar r, r)}}~=~b_1 b_3^{-1}~{\ket { \chi(R;~\bar r,r,r, \bar
r)}}
$$ 
Since
$b_1$ and $b_3$ commute and are diagonal in the same basis with  
same eigenvalues (\ref {eva}), $b_1 b^{-1}_3 {\ket {\phi_l^{side}(\bar
r,r,r,\bar r)}}$ $ = {\ket {\phi_l^{side}(r, \bar r, \bar r,r)}}$.
Thus  the Chern-Simons functional integral for the manifold of Figure (iii) 
is same as that for the manifold in Figure (ii): 
\begin{equation} 
{\ket { \chi(\gamma_1 R;~r,\bar r,\bar r, r)}}~=~{\ket { \chi(R;~r, \bar
r, \bar r, r)}}. \label{mut1}
\end{equation}
Such statements will not hold if we
increase the number of Wilson lines in these manifolds.
 
In order to obtain the action of a $\gamma_2$-mutation, 
let us now consider the Chern-Simons functional integral
over the three-ball shown in Figure (iv).
Notice that this diagarm can be obtained from that in Figure (ii) by
applying ~$b_1b_2b_1b_3b_2b_1$~ on it:
$$
{\ket { \chi(\gamma_2 R;~r,\bar r,\bar r, r)}}~=~b_1b_2b_1b_3b_2b_1~{\ket
{ \chi(R;~r,\bar
r, \bar r, r)}}
$$
\noindent Next we use the fact (Eqns. (\ref {eva}) and (\ref {eva1}))that 
~$b_1~=~b_3$~ in the Hilbert space associated with four-punctured $S^2$
carrying representations $(\bar r, ~\bar r, ~r, ~r)$: 
$$
b_1~{\ket {\phi_l^{side}(\bar r, \bar r,r, r)}}~=~b_3~{\ket
{\phi_l^{side}(\bar r,\bar r,r, r)}} 
$$
\noindent Further,
for an $n$-strand braid on $S^2$, there is an identity
$b_1b_2....b_{n-2}b_{n-1}^2b_{n-2}.....b_2b_1~=~1$.
This in our case $n=4$, reduces to $b_1b_2b_3^2b_2b_1~=~1$.
This
makes the functional integral over the three-ball of Figure (iv)  equal to 
that of Figure (ii):
\begin{equation}
{\ket { \chi(\gamma_2 R;~r,\bar r,\bar r, r)}}~=~{\ket { \chi(R;~r,\bar r, 
\bar r, r)}}
.\label{mut2}
\end{equation}
 
Now the Chern-Simons functional integrals over $S^3$
containing links
$L_1$, $L_2$ and $L_3$,~ $V_r[L_1]$, $V_r[L_2]$ and $V_r[L_3]$,
are given by the products of the dual vector 
$\bra {\chi (S;~\bar r, r,r, \bar r)}$ representing
the functional integral over the manifold shown in Figure (i)
containing room ~\epsfbox{Sec3fig1.eps}~ with 
$\ket {\chi(R; ~r, \bar r, \bar r, r)}$,  
$\ket {\chi(\gamma_1 R; ~r, \bar r, \bar r, r)}$ and 
$\ket {\chi(\gamma_2 R; ~r, \bar r, \bar r, r)}$ 
representing respectively the  fuctional integrals over 
three-balls in Figures (ii), (iii) and (iv):
\begin{eqnarray}
V_r(L_1)&~=~&{\norm {\chi (S;~\bar r, r,r, \bar r)}{\chi(R; ~r, \bar r, \bar
r, r)}}, \nonumber \\
V_r(L_2)&~=~&{\norm {\chi (S;~\bar r, r,r, \bar r)}{\chi(\gamma_1 R; ~r, \bar r,
\bar r, r)}},~ \nonumber \\
V_r(L_3)&~=~&{\norm {\chi (S;~\bar r, r,r, \bar r)}{\chi(\gamma_2 R; ~r, 
\bar r, \bar r, r)}}. \nonumber 
\end{eqnarray}
Equations (\ref {mut1}) and (\ref {mut2}) then imply:
\begin{equation}
V_r[L_1]~=~V_r[L_2]~=~V_r[L_3].
\end{equation}
 
{\it {Thus we have shown that invariants of a link and its
mutants are
identical for every
representation $r$
of a compact semi-simple gauge group, placed on all the
Wilson
lines constituting the links.}}

The well-known invariants viz., Jones\cite{jon} ,
HOMFLY \cite{hom} and Kauffman \cite{kauf}
polynomials are obtained from
$SU(2)$, $SU(N)$ and $SO(N)$ Chern-Simons
theories respectively.
Also  Akutsu-Wadati
polynomials \cite {wad} obtained from $N$ state vertex
models correspond to
$SU(2)$ with spin $N/2$ representation being placed on the
knot/link.
Hence the fact that all these polynomials do not distinguish
mutants is
a special case of the above result.

Now let us give an example of a pair of sixteen crossing mutant knots: 

\vskip0.2cm
\centerline{\epsfbox{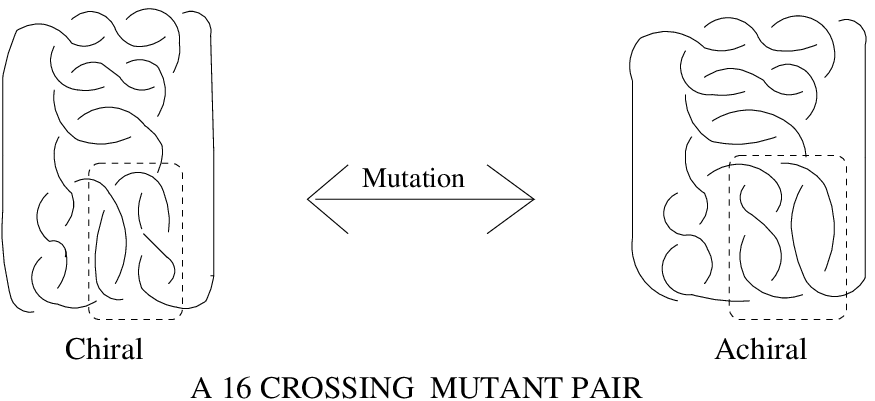}}
\vskip0.2cm

\noindent  The two knots are related by mutation of 
the room indicated by dashed enclosure. One of them 
is chiral, other is not. Since their invariants from 
Chern-Simons theories are the same, here is an example of 
a chiral knot whose chirality can not be detected by any of 
these invariants.

\vskip0.5cm

\section{Exactly solvable  vertex models} The
knot invariants obtained from the Chern-Simons field
theories can also be obtained from statistical mechanical
models in two dimensions\cite{wad,tur,jim} . In these models the 
variables live on the bonds of a square lattice. Their 
properties are described by
the so called $R$- matrix: $R^{m'_1 m'_2}_{m_1 m_2} (u)$,
where $u$ is the spectral parameter. This matrix satisfies 
the Yang-Baxter equation.
Some times another $R$-matrix related to this  by a permutation
is used: $\hat R ~=~ \sigma R $ where the
operation $\sigma$ interchanges the lower two indices: 
~$ {\hat R}^{m'_1 m'_2}_{m_1 m_2} ~=~ R^{m'_1
m'_2}_{m_2 m_1}$. 

The simplest model of interest  is the six-vertex
model of Lieb and Wu \cite{lieb} . The $R$-matrix of this model 
is a $4 \times 4$ matrix with six non-zero enteries. Its 
elements  
$ R^{m'_1~ m'_2}_{m_1~ m_2} (u)$ are explicitly given by:

\begin{center}
\begin{tabular}{c|ccccc}  

$~~~~~~~~~~~~\backslash ({m}'_1 ~{m}'_2)$&$(~{1\over 2}~~~~{1\over 2})$&
$(~{1\over 2}~-{1\over 2})$&$(-{1\over 2}~~~{1\over 2})$&$(-{1\over 2}-{1\over 2})$ \\ 
$({m}_1 ~{m}_2) \backslash~~~~~$ & & & & & \\ 
\hline

$(~{1\over 2}~~~~~{1\over 2})~~~~~$&$sinh(\mu-u)$&0&0&0\\ 
$(~{1\over 2}~-{1\over 2})~~~~~$&0&$-sinhu$&$e^{-u} sinh\mu$&0\\ 
$(-{1\over 2}~ ~~{1\over 2})~~~~~$&0&$e^{u} sinh\mu$&$-sinhu$&0\\ 
$(-{1\over 2}-{1\over 2})~~~~$&0&0&0&$sinh(\mu-u)$\\ 
\end{tabular}
\end{center}

\noindent This may be compactly rewritten as:
\begin{equation}
\epsfbox{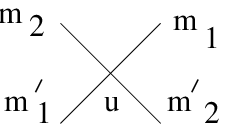}~:~~~R^{m'_1 m'_2}_{m_1 m_2} (u) ~ =~ {\sum_{j~m}}~
\left [ \matrix{{1\over 2}& {1\over 2}& j\cr m_2 &m_1& m} 
\right ]~\lambda_j(u)
~\left [\matrix{{1\over 2}& {1\over 2}& j\cr m'_1 &m'_2& m}
\right ]\label{rm} 
\end{equation}
\noindent where the square brackets here are the $SU(2)$
quantum Clebsch-Gordon coefficients  with deformation
parameter identified as $q~=~e^{2\mu}$; spin $j$ is to be summed
over values $0,~1$ and $m$ takes the associated $m$
values $0$ and $0, \pm 1$ for the two values of
$j$ respectively; and further 
\begin{equation}
\lambda_0 (u) ~= ~ sinh(\mu+u),~~~~ \lambda_1 (u) ~=~ sinh(\mu
- u).
\end{equation}
\noindent The $\hat R$-matrix for this case then reads (same
as above with interchange of $m_1$ and $m_2$ in the
right hand side):
$$
{\hat R}^{m'_1 m'_2}_{m_1 m_2} (u) ~ =~ {\sum_{j~m}}~
\left [ \matrix{{1\over 2}& {1\over 2}& j\cr m_1 &m_2&
m}
\right ]~\lambda_j(u)
~\left [\matrix{{1\over 2}& {1\over 2}& j\cr m'_1
&m'_2& m}
\right ].      
$$ 
\noindent $\lambda_0$ and $\lambda_1$ are the two independent
eigenvalues of $\hat R$. As the spectral
parameter $u$ is taken to infinity, these two
eigenvalues are seen to be proportional to the braiding
eigenvalues ${\lambda_j} ({1\over 2} ~ {1\over 2}) $ for strands
carrying spin half representations  of the $SU(2)_k$ Wess-Zumino
 conformal field theory with the identification $q~=~exp({{2\pi i}\over {k+2}})$.
It is this relation of the $R$-matrix in the limit
$u \rightarrow \infty $ with braiding matrix which allows construction of knot invariants
from the vertex model \cite{wad} . On the other hand,
corresponding to the the general braiding matrices
with braiding eigenvalues ${\lambda_j} (j_1, j_2)$
associated with two strands carrying arbitrary $SU(2)$
representations of spins $j_1$ and $j_2$ obtained from
the conformal field theory, there should be general
solutions of the Yang-Baxter solutions whose
eigenvalues in the limit of large $u$ reduce to these
braiding eigen-values. We propose a
generalization of the formula (\ref{rm}):
\begin{eqnarray}
&\epsfbox{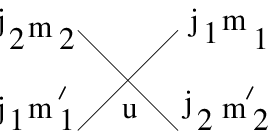}: &\left ( R^{j_1j_2} \right )^{m'_1 m'_2}_{m_1 m_2} (u) ~ =~{\sum_{j~m}}~
\left [ \matrix{j_2& j_1 & j\cr m_2 &m_1&
m}
\right ]\lambda_j(u)
\left [\matrix{j_1 & j_2 & j\cr m'_1
&m'_2& m} \right ]~~~~\label{grm}  \\
&{\rm and}~ &\left ( {\hat R}^{j_1j_2} \right )^{m'_1 m'_2}_{m_1 m_2} (u)
~ =~{\sum_{j~m}}~
\left [ \matrix{j_1& j_2 & j\cr m_1 &m_2&
m}
\right ]\lambda_j(u)
\left [\matrix{j_1 & j_2 & j\cr m'_1
&m'_2& m}
\right ]. \nonumber
\end{eqnarray}
\noindent Here the $SU(2)$ spin $j$ runs over the values allowed for
the irreducible representations  in the product of representations
with spins $j_1$ and $j_2$:~ $j~=~ j_1+j_2,~j_1+j_2-1,~.....~|j_1-j_2|$ ~and 
$m= -j, ~-j+1,~....~j-1, ~j$;~ $m_1, ~m'_1~=~ -j_1,~-j_1+1,~....~j_1$ and 
$m_2, ~m'_2~=$ $~ -j_2,~-j_2+1,~....~j_2$. Only non-zero matrix elements are those with $m~=~m_1+m_2~=~
m'_1+m'_2$. The generalized eigen-values are now given by:
\begin{equation} 
\lambda_j (u)~= \left (\prod_{\ell =|j_1 -j_2 |+1, |j_1 -j_2|+2,...j} 
sinh(\ell \mu-u) \right)\left(\prod_{k=j+1,j+2,...j_1+j_2}sinh(k\mu+u) 
\right ).~~~~~~~\label{geva}
\end{equation}
\noindent In the limit $u \rightarrow \infty $, these  eigen-values do indeed, upto a
proportionality constant, reduce to the braiding
eigenvalues for two strands carrying spins $j_1$ and $j_2$
in the $SU(2)$ conformal field theory.  The generalized $R$-matrix 
satisfies the Yang-Baxter equation:
\begin{eqnarray}
{\sum_{m'_1 m'_2 m'_3}} {\left (R^{j_1 j_2} \right )}^{m'_1 m'_2}_{m_1 m_2} 
(u) {\left (R^{j_1 j_3} \right )}^{m''_1 m'_3}_{m'_1 m_3} (u+v)
{\left (R^{j_2 j_3} \right )}^{m''_2 m''_3}_{m'_2
m'_3}
(v)~~~~~~~~~~~~~~~~~~~~~~~~~~~~~~~~ \nonumber\\
~~~~~~~~~~~~~= ~{\sum_{m'_1 m'_2 m'_3}} {\left (R^{j_2 j_3}
\right )}^{m'_2 m'_3}_{m_2
m_3} (v)
{\left (R^{j_1 j_3} \right )}^{m'_1 m''_3}_{m_1 m'_3} (u+v) 
{\left (R^{j_1 j_3} \right )}^{m''_1 m''_2}_{m'_1
m'_2}(v)~.
~~~~~~~~~\label{gyb}
\end{eqnarray}

For $j_1 =j_2 = {1\over 2}$, this
$R$-matrix is the same as that of $6$-vertex model
above. For other low values of  $(j_1 , j_2) $, these
solutions of the Yang-Baxter equation correspond to
other well known vertex models:
\begin{center}
\begin{tabular}{cc}
$\left ({R^{{1\over 2}~{1\over 2}}} \right )^{m'_1 m'_2}_{m_1 m_2} (u)$:~~~& ~~~~~~$6$-vertex model of Lieb and Wu\cite{lieb} \\
$\left ({R^{1~1}} \right )^{m'_1 m'_2}_{m_1 m_2} (u)$:~~&~~~~~~$19$-vertex model of Zamolodchikov and Fateev\cite{zam}\\
$\left ({R^{{3\over 2}~{3\over 2}}} \right )^{m'_1 m'_2}_{m_1 m_2} (u)$:~~& ~~~~~~$44$-vertex model\\
$\left ({R^{2~2}} \right )^{m'_1 m'_2}_{m_1 m_2} (u)$:~~& ~~~~~~~$85$-vertex model\\
......~~&~~~~~~......\\
......~~&~~~~~~...... 
 
\end{tabular}
\end{center}

Thus, an alternate route to  the same knot
invariants as those emerge from the Chern-Simons
theory is to obtain the braid representations from the
$\hat R$-matrix by taking the limit $u \rightarrow \infty$:

\centerline{\epsfbox{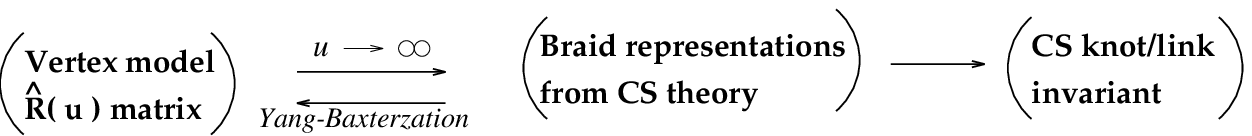}}

\vskip0.5cm

\section{Three-manifold invariants} The invariants of knots and
links in $S^3$ obtained from the Chern-Simons theory can be used 
to construct three-manifold invariants\cite{wit,res,lic} . This 
provides an important tool to study  topological  properties of 
three-manifolds.  Starting step in this construction is a theorem due 
to Lickorish and Wallace \cite{wal,rol}:
\vskip0.2cm

{\bf Fundamental theorem of Lickorish and Wallace:}~{\it  Every
closed, orientable, connected three-manifold, $M^3$ can be
obtained by surgery on an unoriented framed knot or link $[L, ~f]$ 
in $S^3$.}
\vskip0.2cm

The framing $f$ of a link  $L$ is defined by associating 
with every component knot $K_s$ of the link an
accompanying closed curve $K_{sf}$ parallel to the knot and
winding $n(s)$ times in the right-handed direction. That is
the linking number $lk(K_s, K_{sf})$ of the component knot 
and its frame is $n(s)$. A particular framing is the so
called vertical framing where the frame is thought to be
just vertically above the two dimensional projection of the
knot as shown below. We may indicate this sometimes by 
putting $n(s)$ writhes in the strand making the knot or even
by just simply writing the integer $n(s)$ next to the knot
as shown below: 

\centerline{\epsfbox{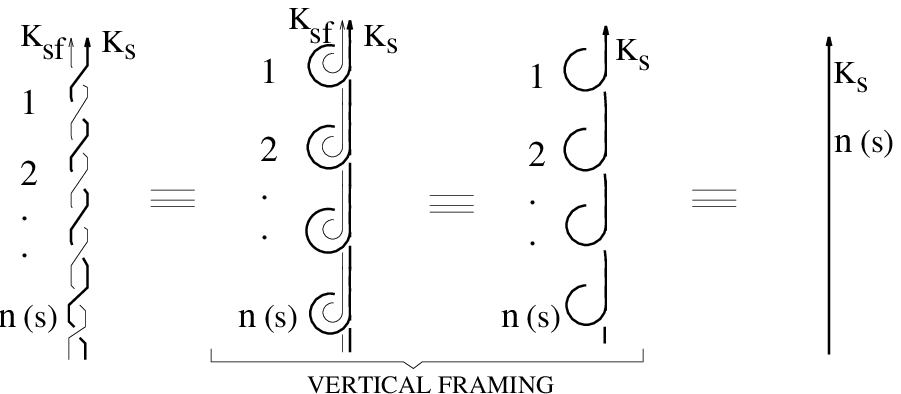}} 
\vskip0.2cm

Next the surgery on a framed link $[L, f]$  made of
component knots $K_1, K_2, ....~K_r$  with framing
$f~=~ (n(1), n(2), ....~n(r))$ in $S^3$ is performed in the following
manner. Remove a small open solid torus neighbourhood ${N_s}$
of each component knot $K_s$, disjoint from all other such
open tubular neighbourhoods associated with other component
knots. In the manifold left behind $S^3 -(N_1 \cup N_2
\cup~....~N_r)$, there are $r$ toral boundaries. On each such
boundary, consider a simple closed curve (the frame) going 
$n(s)$ times along the meridian and once along the longitude 
of the associated knot $K_s$. Now do a modular transformation on
such a toral boundary such that the framing curve bounds a
disc. Glue back the solid tori into the gaps. This yields a
new manifold $M^3$. The theorem of Lickorish and Wallace
assures us that every closed, orientable, connected
three-manifold can be constructed in this way. 

This construction of three-manifolds  by surgery is not unique: 
surgery on more than one framed link can yield homeomorphic 
manifolds. But the rules of equivalence of framed links in
$S^3$ which yield the same three-manifold on surgery are
known. These rules are known as Kirby moves.
\vskip0.2cm

{\bf Kirby calculus on framed links in $S^3$:}
Following two elementary moves (and their inverses) generate 
Kirby calculus\cite{kir}:

\vskip0.2cm

{\it Move I}. For a number of unlinked strands belonging
to the component knots $K_s$ with framing $n(s)$ going
through an unknotted circle $C$ with framing $+1$, the
unknotted circle can be removed after making a
complete clockwise twist from below in the disc enclosed 
by the circle $C$: 

\centerline{\epsfbox{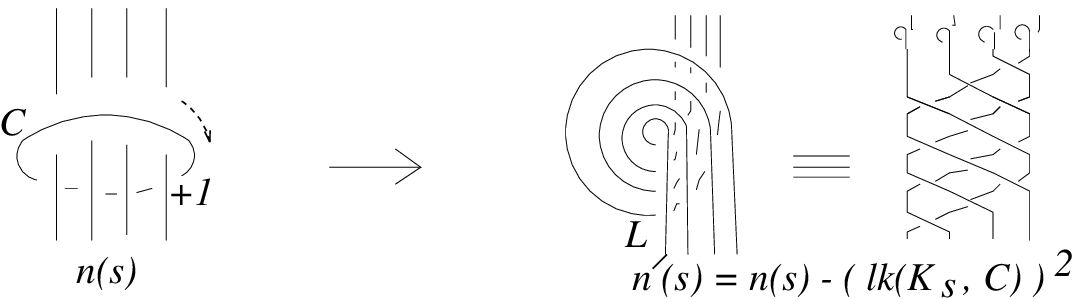}} 
\vskip0.2cm

\noindent  In the process, in addition to
introducing new crossings, the framing of the various
resultant component knots, $K'_s$ to which the affected  strands
belong, change from $n(s)$ to $n'(s)~=~n(s)- \left (
lk(K_s, C) \right )^2$.
\vskip0.2cm

{\it Move II}. Drop a disjoint unknotted circle with
framing $-1$ without any change in the rest of the
link:

\centerline{\epsfbox{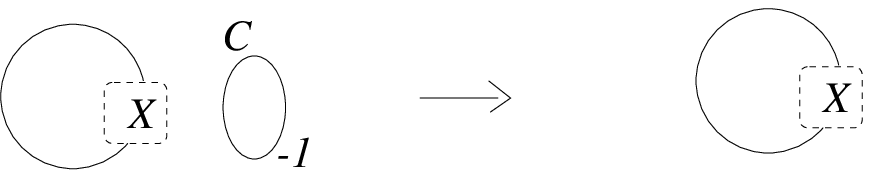}} 
\vskip0.2cm

\par Two Kirby moves (I) and (II) and their inverses
generate the conjugate moves\cite{lic}:
\vskip0.2cm

{\it Move ${\bar I}$}. Here a circle $C$ with framing $-1$
and enclosing a number strands can be removed after
making a complete anti-clockwise twist from below in
the disc bounded by the curve $C$:

\centerline{\epsfbox{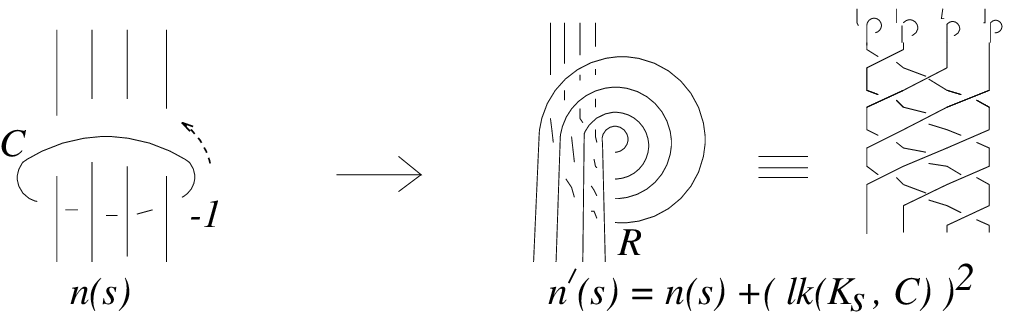}} 
\vskip0.2cm

\noindent Again, this changes the framing of the
resultant knots $K'_s$ to which the enclosed strands
belong from $n(s)$ to $n'(s)~=~n(s) + \left (
lk(K_s, C) \right )^2$.
\vskip0.2cm

{\it Move ${\bar {II}}$}. A disjoint unknotted circle
with framing $+1$ can be dropped without affecting
the rest of the kink:

\centerline{\epsfbox{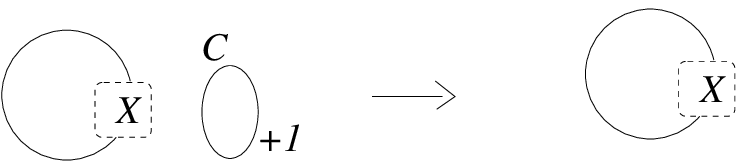}} 
\vskip0.2cm 

\par  Thus Lickorish-Wallace theorem  and equivalence of
surgery under Kirby moves  reduces the theory of closed,
orientable, connected three-manifolds to the theory of
framed unoriented links via a one-to-one correspondence:
$$
\left( \matrix {Framed~ links~ in~ S^3~ modulo \cr
 equivalence~under~ Kirby ~moves} \right) ~\leftrightarrow
~
\left( \matrix
{Closed, ~orientable,~ connected~ three-\cr manifolds~ 
modulo~ homeomorphisms} \right)
$$
\noindent This consequently allows us to characterize 
three-manifolds by the invariants of the associated 
unoriented framed knots and links obtained from the
Chern-Simons theory in $S^3$. This can be done by
constructing an appropriate combination of the invariants
of the framed links which is unchanged under Kirby moves: 
$$
\left( \matrix {Invariants ~of ~a ~framed ~unoriented ~link \cr
 which ~ do~ not~ change ~under~ Kirby ~moves} \right) ~=~~
\left( \matrix
{Invariants ~ of ~associated \cr three-manifold~
} \right)
$$

We shall now construct one such invariant from the link invariants of 
$SU(2)$  Chern-Simons theory.

\vskip0.5cm

{\bf Invariants for unoriented knots and
links from $SU(2)$ Chern-Simons theory:}~ The knot/link invariants we discussed earlier were
constructed for oriented links with standard framing. The
braiding eigenvalues given in Sec. 2 reflect this
property. For our present purpose, we need to have invariants for
unoriented links in vertical framing. This is achieved by
taking the eigen-values for the braid matrix introducing
right-handed or left-handed $(R/L)$ half-twist in two parallel 
strands carrying spins $j, ~j'$ as:
\begin{eqnarray}
\epsfbox{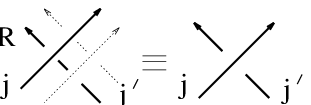}:~~~~  &\lambda^{(+)}_{\ell ,R} (j,~j')~ =~ \lambda_{\ell}(j,~j') ~=~
(-)^{|j-j'|-\ell} q^{-(C_j +C_{j'} -C_{\ell} )/2}, \nonumber\\
\epsfbox{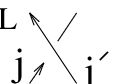}:~~~~  &\lambda^{(+)}_{\ell ,L} (j,~j')~ =~ (\lambda_{\ell}(j,~j')
)^{-1} ~=~
(-)^{|j-j'|-\ell} q^{(C_j +C_{j'} -C_{\ell} )/2}, \nonumber 
\end{eqnarray}
\noindent and for anti-parallel strands:
$$ 
\epsfbox{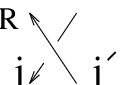}:~~~  \lambda^{(-)}_{\ell ,R} (j, ~j')~ =~ (\lambda_{\ell}(j,~j')
)^{-1},~~~~~~~~~~~
\epsfbox{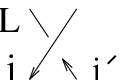}:~~~  \lambda^{(-)}_{\ell ,L} (j,~j')~ =~ \lambda_{\ell}(j,~j').
$$   
\noindent Clearly these eigenvalues do not see the
orientations on the strands; these are sensitive only to
over-crossing and under-crossing.

In standard framing, a writhe can be stretched without affecting the
link. In vertical framing this is not so. In this case the invariants 
of knots get changed by a phase when a writhe is smoothed out as: 
$$
\epsfbox{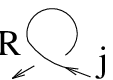} =~ \left(\lambda_0 (j,~j)\right)^{-1}~\epsfbox{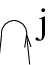}
=~q^{C_j}~\epsfbox{Sec5fig11.eps},
~~~~~{\rm and}~~~\epsfbox{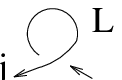}=~ \lambda_0
(j,~j)~ \epsfbox{Sec5fig11.eps}
=~  q^{-C_j} ~\epsfbox{Sec5fig11.eps}
$$
\noindent Thus, invariant for an unknot with framing $+1$ and $-1$ is
related to the invariant for an unknot with zero framing as:
$$
\epsfbox{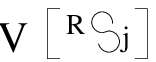} =~ q^{C_j} ~\epsfbox{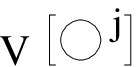} =~q^{C_j}
[2j+1], ~~~{\rm and}~~ \epsfbox{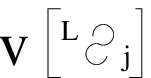}=~ q^{-C_j}
\epsfbox{Sec5fig13.eps}=~q^{-C_j} [2j+1] 
$$ 

The invariant for a Hopf link carrying spins $j_1$ and $j_2$
on the component knots and with vertical framing can be obtained in
two ways using the braiding and inverse braiding:
\begin{eqnarray}
 \epsfbox{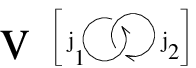} ~&=&~ \sum_{\ell}~[2\ell +1]~\left (
\lambda_{\ell} (j_1,
~j_2) \right )^2
~~= ~q^{-C_{j_1} -C_{j_2}} ~ \sum_{\ell}~ [2\ell
+1]~q^{C_{\ell}}, \nonumber\\
\epsfbox{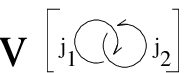}  ~&=&~ \sum_{\ell}~[2\ell +1]~\left ( \lambda_{\ell} (j_1,
~j_2) \right )^{-2}  
~~= ~q^{C_{j_1} +C_{j_2}} ~ \sum_{\ell}~ [2\ell
+1]~q^{-C_{\ell}}.\nonumber  
\end{eqnarray}
\noindent These are equal, as they should be, due to the
identity:
$$
q^{-C_{j_1} -C_{j_2}} ~ \sum_{\ell} ~[2\ell
+1]~q^{C_{\ell}} ~= ~q^{C_{j_1} +C_{j_2}} ~ \sum_{\ell}
~[2\ell 
+1]~q^{-C_{\ell}}~=~ [(2j_1 +1)~(2j_2 +1)].
$$
\noindent Consider  next  the Hopf link $H(j_1,~j_2)$ with  framing $+1$
 for each of its component knots: 
\vskip0.2cm 
\centerline{\epsfbox{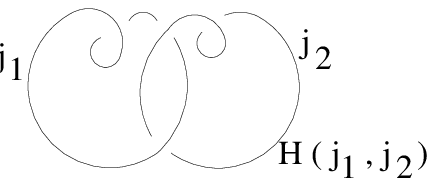}}
\vskip0.2cm

\noindent The invariant for this link is given by 
\begin{equation} 
V[H(j_1,~j_2)] ~=~ q^{C_{j_1} +C_{j_2}} ~~ \epsfbox{Sec5fig16.eps}
~=~ q^{C_{j_1}
+C_{j_2}}~ [(2j_1 +1)~(2j_2 +1)].
\end{equation}

Next we wish to construct a combination of these invariants
which would be unchanged under Kirby move I:

\vskip0.2cm                                              
\centerline{\epsfbox{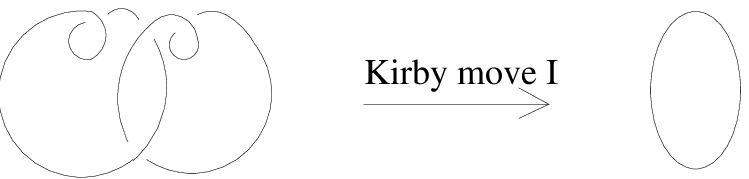}}
\vskip0.2cm

\noindent That is, we
solve the  following equation  for $\mu_{\ell}$ and
$\alpha$: 
\begin{equation}
\sum_{\ell=0, 1/2, 1,...~k/2} ~\mu_{\ell}~V[ H(j,~\ell)] ~=~ \a ~[2j
+ 1]~, 
\end{equation} 
\noindent where $[2j+1]$ is the invariant for an unknot
carrying spin $j$ representation.  This solution is
given by
\begin{equation}
\mu_{\ell} ~=~ S_{0 \ell}~, ~~~~~~~~~\a ~=~ e^{\pi ic/4}~,
~~~~~~~~c~=~ {3k\over{k+2}}~, 
\end{equation}
$$ {\rm where}~~~~~~~~~~~~~~~~~
S_{j \ell} ~= ~ \sqrt{2\over {k+2}}~~ sin~ {{\pi (2j+1)~(2\ell
+1)}\over {k+2}}~ =~ [(2j+1)~(2\ell +1)]~S_{0 0}~.
$$
\noindent This can be easily verified by using the
identity: 
$$
\sum_{\ell}~ S_{j \ell} ~q^{C_{\ell}}~S_{\ell m} ~=~
e^{\pi ic/4}~ q^{-C_j -C_m}~S_{j m}~,
$$
which follows readily by noticing that the
matrices $S_{j \ell}$ and $T_{j \ell}$$~=~q^{C_j}~
e^{-\pi ic/12}~\delta_{j\ell}$ are the generators
of the modular transformations $\tau \rightarrow
-{1/\tau}$ and $\tau \rightarrow \tau +1$ on the
characters of the Wess-Zumino $SU(2)_k$ conformal
field theory and hence satisfy  relations $S^2 ~=~1$,
and $(ST)^3~=~1$. This last relation implies
$STS~=T^*S~T^*$ which is the identity above.

Now let us consider the following two links $ H(X;~ j,~\ell)$  and 
$U(X;~j)$: 

\vskip0.2cm 
\centerline{\epsfbox{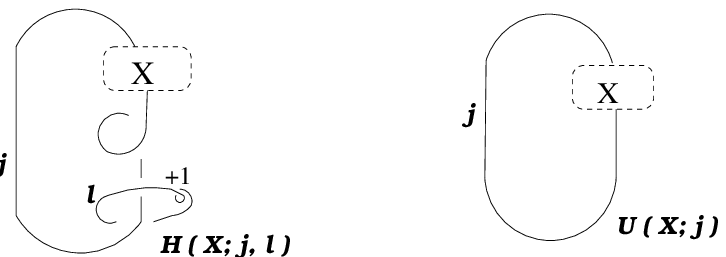}}
\vskip0.2cm 

\noindent where  $X$ as an arbitray entanglement inside the
box. The link $H(X;~j,~\ell)$ is the  connected sum of the link $U(X;~ j)$
and a framed Hopf link $H(j,~\ell)$. Factorization properies of invariants of
such a connected sum of links yields:  
$$
[2j+1]~V[H(X;~j,~\ell)]~=~V[U(X;~j)]~~V[H(j,~\ell)]~.
$$
\noindent This further implies:
$$
\sum_{\ell}~ \mu_{\ell}~V[H(X;~j,~\ell)] ~=~ \alpha~V[U(X;~j)]~.
$$
\noindent It is possible to generalize this  relation
for the following links $H(X;~j_1,~j_2,~....~j_n;~\ell)$ and $U(X;~j_1,~j_2,~....~j_n)$:
\vskip0.2cm
\centerline{\epsfbox{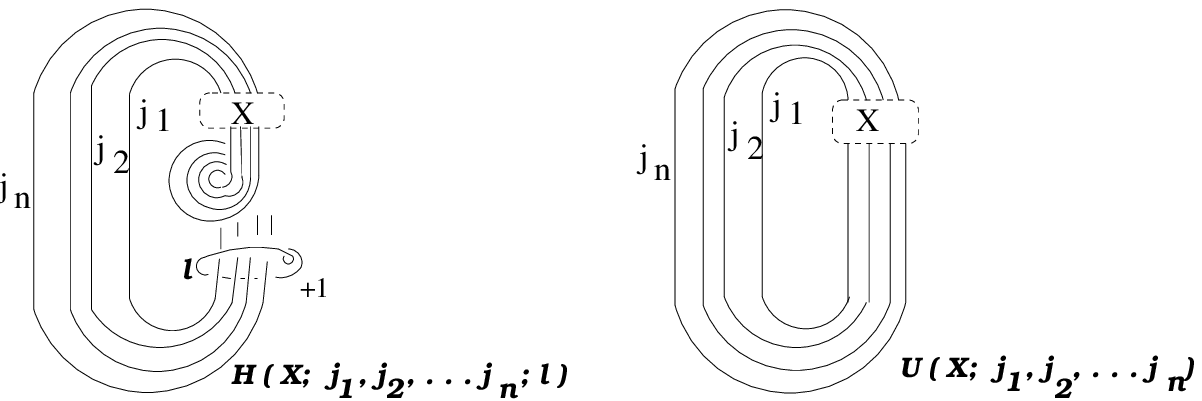}}
\vskip0.2cm
\noindent This relation then reads:
\begin{equation}
\sum_{\ell}~ \mu_{\ell}~V[H(X;~j_1,~j_2,...~j_n; ~\ell)] ~=~ 
\alpha~V[U(X;~j_1,~j_2, ....~j_n)]~.\label{kir1} 
\end{equation}

 Also, for a link containing a disjoint unknot with framing
$-1$: 
\begin{equation}                                           
\sum_{\ell}~ \mu_{\ell}~\epsfbox{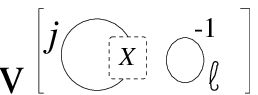}  ~=~
\alpha^* ~\epsfbox{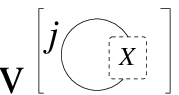} \label{kir2}
\end{equation}
\noindent This follows readily due to the exact
factorizations of invariants of  disjoint links into those
of the  individual links and use of the identity:
~~$\sum_{\ell}~ S_{0 \ell} ~q^{-C_{\ell}}~S_{\ell 0} ~=~
e^{-\pi ic/4}~ S_{0 0}$.

Clearly the Eqns. (\ref{kir1}) and (\ref {kir2}) 
respectively correspond to the two generators of the Kirby
calculus. Therefore, this allows us to state
the following result:

\vskip0.2cm

{\it For a framed unoriented link $[L,~f]$ with
component knots $K_1,~K_2, ....~K_r$  and framing
$f~=~(n_1,~n_2,....~n_r)$, the following is an invariant
of the associated closed, connected, orientable
three-manifold obtained by surgery on the link (upto
possible changes of the framing of the manifold): }
\begin{equation}
F[L,~f]~=~ \sum_{j_i=0,1/2,1,....k/2}~\mu_{j_1}~\mu_{j_2}~....~\mu_{j_r}~ V[L;~
n_1,~n_2,....~n_r;~ j_1,~j_2,....~j_r]
\end{equation}
\par
Under Kirby moves $F[L,~f]$ changes only by possible phase
factors ({\it i.e.}, powers of $\alpha$ or $\alpha^*$)
associated with the framing of the manifold.
Particularly, under the two Kirby generators $F[L;~f]$
changes as:
	
\vskip0.2cm 
\centerline{\epsfbox{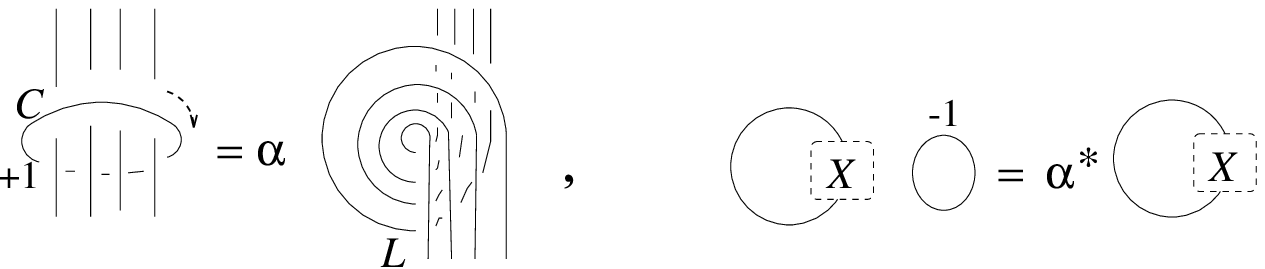}}
\vskip0.2cm

\noindent  Here we have depicted the manifold invariant $F$ 
by the affected portion of the diagram of the framed link on
which surgery is to be performed.

The frame dependence of the manifold invariant $F$ can be
compensated  by noticing the following property of the linking
matrix. For a framed link $[L,~f]$ whose component knots
$K_1,~K_2,~....~K_r$ have framings (self-linking numbers) as $n_1,~n_2,~....~n_r$
respectively, the linking matrix is defined as 
$$
W[L,~f]~=~\left ( \matrix { n_1 &lk(K_1,K_2) &lk(K_1,K_3)
&.....&lk(K_1,K_r) \cr
lk(K_2,K_1) & n_2& lk(K_2,K_3)&.....&lk(K_2,K_r) \cr
..&..&n_3&.....&.. \cr
..&..&..&.....&..\cr
lk(K_r,K_1)&..&..&.....&n_r} \right )
$$
\noindent where $lk(K_i,K_j)$ is the linking number of knots $K_i$
and $K_j$. The signature of the linking matrix is given by
$$
\sigma [L,~f]~=~({\rm no. ~of ~+ve ~eigenvalues~of}~ W)-(
{\rm no.~of~-ve~eigenvalues~of}~W)
$$
\noindent Then this signature for the framed link $[L,~f]$
and those for the links $[L',~f']$ obtained by
transformation by  the two elementary
generators of the Kirby calculus are related in a simple
fashion:
$$
Kirby~ move~ I:~\sigma [L,~f]~=~ \sigma ~[L',~f']+1~;~~~
~~~~Kirby~move~II:~\sigma [L,~f] ~=~\sigma ~[L',~f']-1~.
$$

Now, if we define a new three-manifold invariant by
$\hat F[L,~f]~=~\alpha^{-\sigma[L,~f]}~F[L,~f]$,
then this invariant will not see the changes of
manifold framings under Kirby moves; it would be
exactly unchanged by the Kirby moves. Thus we may
state the following important result:

\vskip0.2cm

{\bf Proposition:}~{\it For a framed link $[L,~f]$ with
component knots, $K_1,~K_2,~....~K_r$ and their framings
respectively as $n_1,~n_2,~....~n_r$, the quantity}
\begin{equation}
{\hat F[L,~f]}~=~ \alpha^{-\sigma[L,~f]}~
\sum_{\{j_i\}}~~\mu_{j_1}~\mu_{j_2}~....~\mu_{j_r}~ V[L;~n_1,~n_2,~...~n_r;~ 
j_1,~j_2,~....~j_r]
\end{equation}
\noindent {\it constructed from invariants $V$ of the unoriented
framed link, is an invariant of the associated three-manifold
obtained by surgery on that link.} 

\vskip0.2cm

{\bf Explicit  examples:} ~Now let us give the value of this
invariant for some simple three-manifolds. The
surgery descriptions of manifolds $S^3$, $S^2 \times S^1$ and
$RP^3$ are given by an unknot with framing $+1,~0$ and $+2$
respectively. The  above invariant for these manifolds is:
$$
{\hat F}[S^3]~=~1~, ~~~~~{\hat F}[S^2 \times S^1]~=~ {1 \over
S_{00}}~,~~~~~~{\hat F}[RP^3]~=~
\alpha\sum_{j=0,{1\over 2},1,...{k\over 2}}{S_{0j} ~q^{-2C_j}~ S_{j0}
\over S_{00}}~.
$$

A more general example is the whole class of Lens spaces
${\cal L}(p,~q)$; above three manifolds are special
cases of this class of manifolds. These are obtained\cite{rol}
by surgery on a framed link made of
successively linked unknots with framing given by integers
$a_1,~a_2,~......~a_n$:

\centerline{~\epsfbox{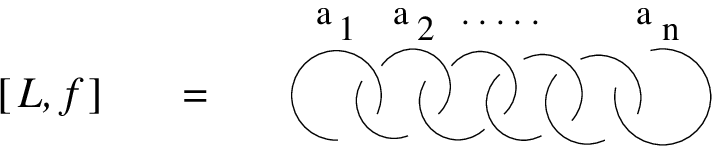}}
\vskip0.2cm

\noindent 
where these framing integers provide a continued fraction
representation for the ratio of two integers $p,~q$:
$$
{p \over q} ~=~ a_n -{1 \over{a_{n-1} -{1 \over{.......~a_3- {1
\over {a_2 -{1 \over a_1}}}}}}}~.
$$
\noindent The invariant for these manifolds can readily be evaluated
and is given by the formula:
$$
{\hat F}[{\cal L}(p,~q)]~=~
{\alpha}^{-\sigma[L,~f]}~{\alpha}^{(\sum a_i )/3}~
~{(S~M^{(p,~q)} )_{00} \over S_{00}}~,
$$
\noindent where  matrix $M^{(p,~q)}$ is given in terms of the modular
matrices $S$ and $T$:
$$ 
M^{(p,~q)}~=~ T^{a_n} S~ T^{a_{n-1}} S~ ......~T^{a_2}
S~ T^{a_1} S~.
$$

Another example we take up is the Poincare manifold
$P^3$ (dodecahedral space). It is given \cite{rol} by surgery
on a right-handed trefoil knot with framing $+1$:

\vskip0.2cm
\centerline{\epsfbox{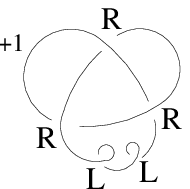}}
\vskip0.2cm
\noindent Notice, each right-handed crossing of the trefoil
introduces $+1$ linking number between the knot and its vertical
framing, and each of the two left-handed writhes contributes $-1$ so
that the total frame number of this knot is $+1$.
Using the proposition above, the invariant  for this
manifold can be calculated. It turns out to be:
$$
{\hat F}[P^3]~=~{\alpha}^{-1}~~\sum_{m,\ell,j=0,{1 \over
2}, 1,~...{k \over 2}}~~{(-)^j {S_{0\ell}~S_{0j}~S_{\ell m}~
S_{0 \ell}~S_{jm}~q^{-5C_{\ell} +{3 \over 2}C_j}}
\over {S_{00}~S_{0m}}}~.
$$
 
It is of interest that this invariant by inspection
turns out to be related in a simple way to the partition 
functions $Z[M^3]$  of the Chern-Simons theory in  
all those three-manifolds $M^3$ for which these have been calculated:
${\hat F}[M^3]~=~Z[M^3]/S_{00}$. This thus provides an
alternative method of calculating the Chern-Simons
partition function.
 
The three-manifold invariants presented here use link invariants from
 $SU(2)$ Chern-Simons theory. It is clear that a similar construction can
be done with link invariants from Chern-Simons gauge theories based on
other semi-simple groups. These would yield new three-manifold invariants.

Next question we may ask is: ~Is this three-manifold
invariant complete? ~Two  manifolds $M$
and $M'$ for which the invariants ${\hat F}[M]$ and ${\hat F}[M']$
are different can not be homeomorphic to each other. 
But the converse is not always true; for  two arbitrary 
manifold, the invariants  need not be always different.
Recall the invariants obtained from
Chern-Simons theory  for mutant knots are not
distinct. Hence, manifold obtained by surgery on mutant
knots can not be distinguished by this three-manifold
invariant.

\end{document}